# Deep Clustering for Mars Rover image datasets

Vikas Ramachandra

Adjunct Professor, Data Science

University of California, Berkeley

Berkeley, CA

virama@berkeley.edu

## Abstract

In this paper, we build autoencoders to learn a latent space from unlabeled image datasets obtained from the Mars rover. Then, once the latent feature space has been learnt, we use k-means to cluster the data. We test the performance of the algorithm on a smaller labeled dataset, and report good accuracy and concordance with the ground truth labels. This is the first attempt to use deep learning based unsupervised algorithms to cluster Mars Rover images. This algorithm can be used to augment human annotations for such datasets (which are time consuming) and speed up the generation of ground truth labels for Mars Rover image data, and potentially other planetary and space images.

## 1. Introduction and background

The Curiosity rover captured several sets of images of the landscape on Mars. This image dataset is publicly available [1]. Given the sheer volume of images captured by the rover over multiple years, it is an extremely time consuming task to manually inspect each image and find interesting patterns or objects from the dataset. Computer vision and machine learning algorithms can help with this effort. Previous work in this direction can be summarized as follows.

In [2], the authors have built a supervised learner: a soil and object classification system using the Alexnet CNN. Other work has looked at segmentation, as well as robot navigation tasks for the rover, based on image analysis. [3], [4].

In this paper, we explore unsupervised clustering to help discover interesting groupings of this image data. We use a combination of autoencoders and k-means to achieve our goal, and further details and results are outlined below. To the best of our knowledge, this is the first effort to use unsupervised machine learning to solve this problem.

## 2. Deep clustering algorithm

An autoencoder is an artificial neural network used for unsupervised learning of efficient coding of the input data [7]. The aim of an autoencoder is to learn a representation (encoding) for a set of data, typically for the purpose of dimensionality reduction.

**Deep learning based clustering: Autoencoders**

Architecturally, the simplest form of an autoencoder is a feedforward, non-recurrent neural network very similar to the multilayer perceptron (MLP) – having an input layer, an output layer and one or more hidden layers connecting them – but with the output layer having the same number of nodes as the input layer, and with the purpose of reconstructing its own inputs (instead of predicting the target value). Therefore, autoencoders are unsupervised learning models.

An autoencoder always consists of two parts, the encoder and the decoder, which can be defined as transitions, $(\phi, \psi)$ such that:

$\phi : X \rightarrow F$, encoder
$\psi : F \rightarrow X$, decoder
$(\phi, \psi) : argmin_{(\psi,\phi)} \| X - (\psi * \phi)X \|$, in the L-2 norm sense

The nonlinear functional mappings for the encoder and decoder are learnt to minimize the reconstruction error above. The learned mapping, if it maps the input to a lower dimensional encoding, becomes a form of non-linear dimensionality reduction technique.

The training algorithm for an autoencoder can be summarized as
For each input x,
Do a feed-forward pass to compute activations at all hidden layers, then at the output layer to obtain an output x'
Measure the deviation of x' from the input x (typically using squared error),
Backpropagate the error through the net and perform weight updates.
Repeat the above steps for several epochs until the error reaches below a certain threshold or converges.

**Autoencoders for learning a latent space embedding**

We build an autoencoder with the following structure. If the input data as NXN dimensions (for images), the first and last layers of the autoencoder have NxN neurons, Our aim is to reduce the dimensions, to M, so the middle layers of the autoencoder has M neurons, as shown in the figure below (left) for illustration. The training process will try and learn the weights in an iterative fashion, using mean squared error loss function and gradient backpropagation.

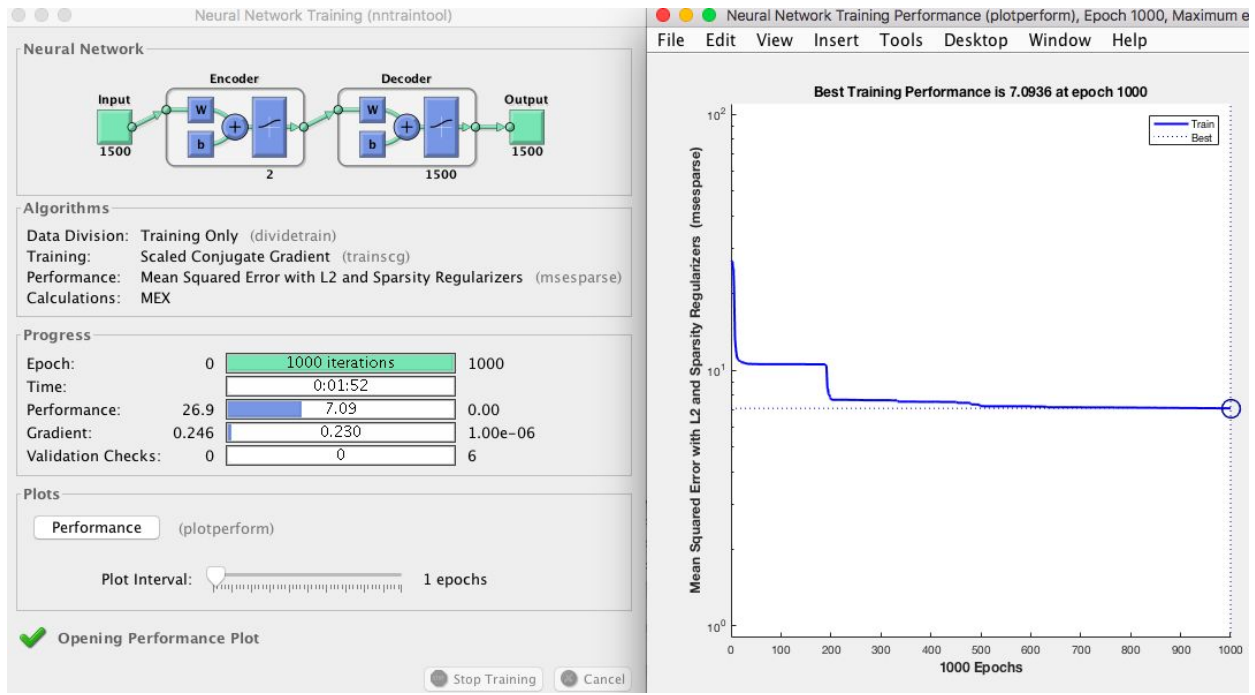

Figure: Left: The autoencoder network, Right: the training mean squared error at each epoch.

**K-means**

Once the autoencoder has been trained on unlabeled data, all images are projected onto the latent space and standard k-means is run to assign data points to their respective clusters.

**Datasets, evaluation metrics and results**

The large unlabeled dataset [5] consists of 32,000 images from the MastCam (color images), and contain timestamps, but no other object or image category labels. For building the autoencoder, we use this large dataset without labels [5]. Then, we use k-means on the learnt latent feature space. For evaluation, we compare the k-means clusters obtained, with labelled object groups in a smaller dataset, which does have ground truth [6]. For this, we project each image from the labeled dataset into the same latent space, and form clusters. The dataset consists of 6691 images spanning 24 classes.

We obtained 92% accuracy using the autoencoder+k-means algorithm, hen we compared the clusters to the ground truth labels for the dataset in [6]. The confusion matrix is visualized below.

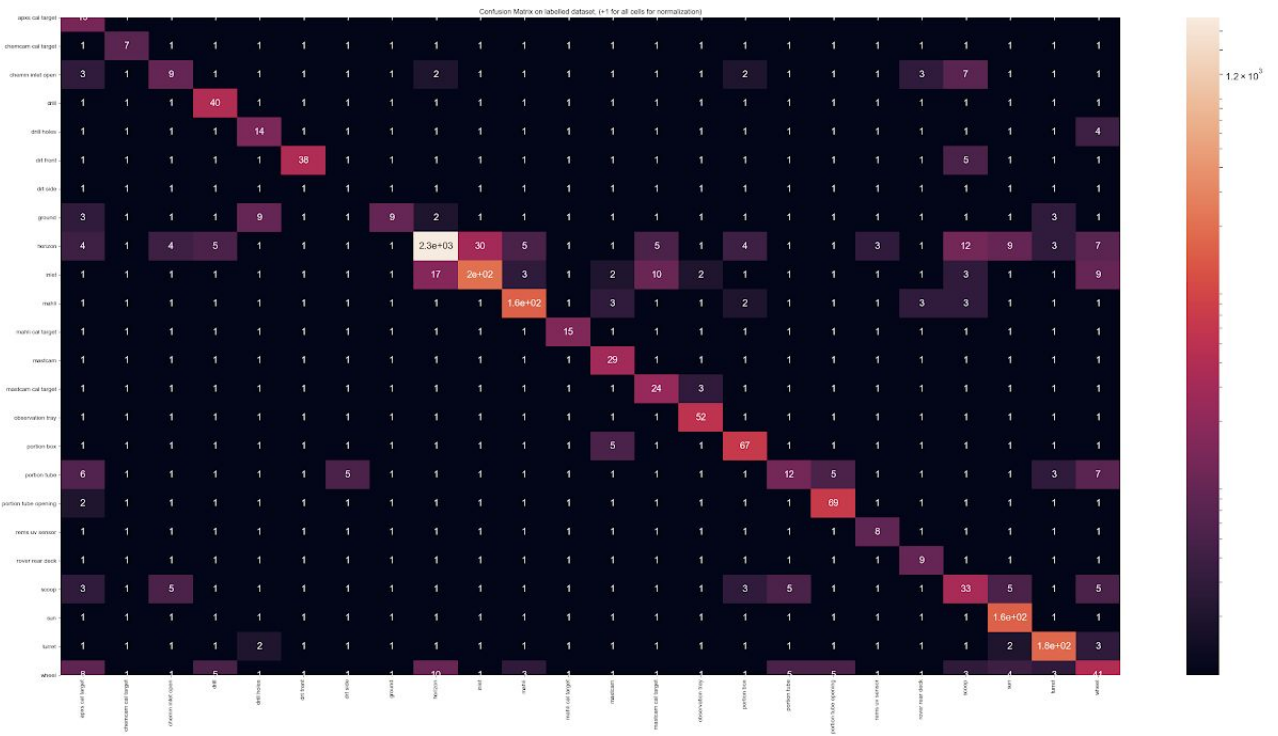

Figure: Confusion matrix for the autoencoder+k-means algorithm, compared to the ground truth labels, on the labeled dataset [6]

Some example results of true class labels, and data points in the matched cluster found by the algorithm.

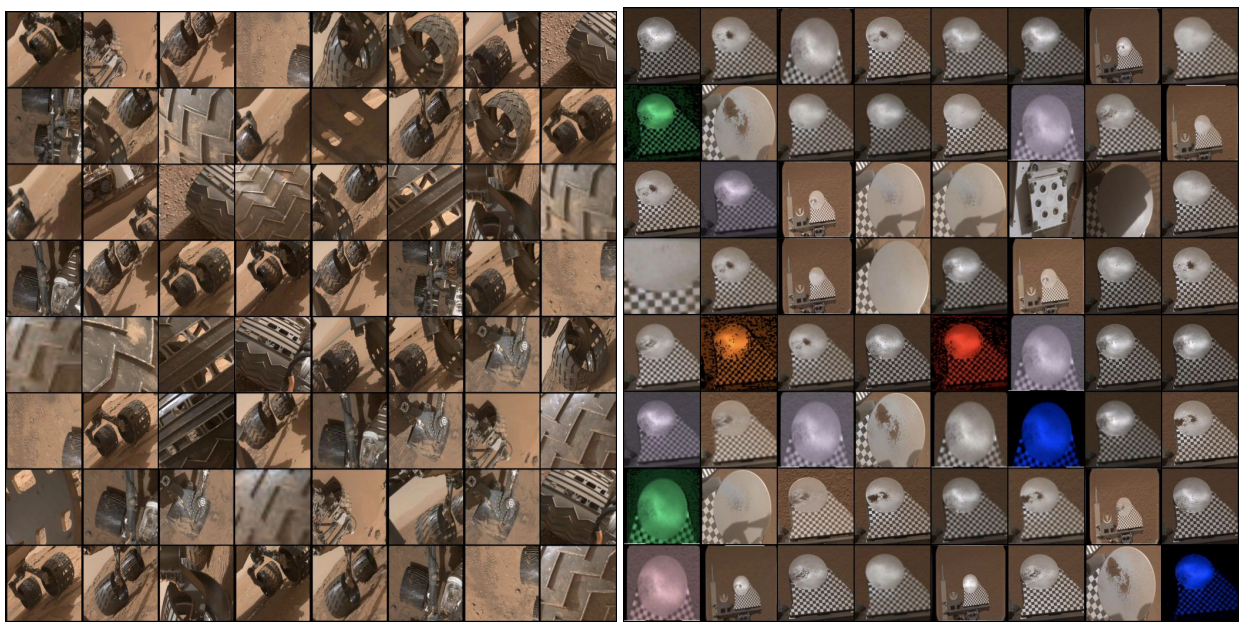

Wheel image cluster                                  Observation tray cluster

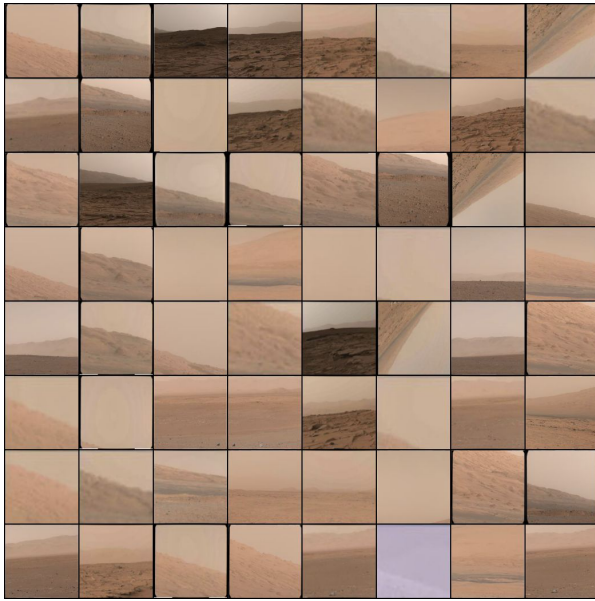 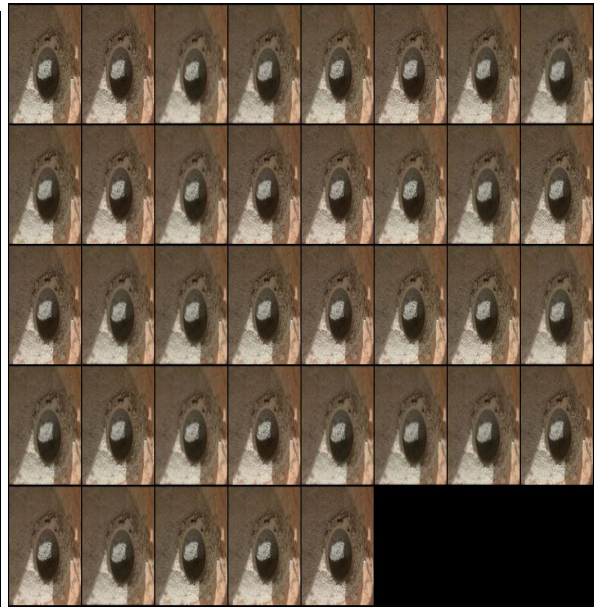

Horizon image cluster          Drill hole image cluster

## 3. Conclusion

We have shown that autoencoders can be used to learn a latent feature space for unlabeled image datasets obtained from the Mars rover. Then, the images are projected onto the learned latent feature space, and k-means is used to cluster the data. We test the performance of the algorithm on a smaller labeled dataset, and report good accuracy and concordance with the ground truth labels. This is the first attempt to use deep learning based unsupervised algorithms to cluster Mars Rover images. This algorithm can be used to augment human annotations for such datasets (which are time consuming) and speed up the generation of ground truth labels for Mars Rover image data, and potentially other planetary and space images.

Our code can be found here:
https://github.com/vikas84bf/MarsPlaySet

## References

[1] Mars Curiosity rover image gallery
 https://www.nasa.gov/mission_pages/msl/images/index.html